%% file: main-ellipjanus.tex
\documentclass[superscriptaddress,twocolumn]{revtex4-1}
\setcitestyle{super,open={[},close={]}}
\usepackage{lipsum}
\usepackage{mathrsfs}
\usepackage{bm,amsbsy,amssymb,amsmath}
\usepackage{caption}
\usepackage{subcaption}
\usepackage{graphics,graphicx,dcolumn,fleqn,epic,eepic,float,tabularx}
\usepackage{multirow,rotate,rotating,color}
\usepackage[utf8]{inputenc}
\newcommand{\figref}[1]{Figure~\ref{fig:#1}}
 
\newcommand{\citeref}[1]{$^[$\cite{#1}$^]$}
  \definecolor{tuered}{RGB}{214,0,74}
  \definecolor{tueblue}{RGB}{0,102,204}
  
  \newcommand{\revisedtext}[1]{\textcolor{black}{#1}}

\usepackage{tikz,pgfplots}
\usepackage{relsize}
\tikzset{fontscale/.style = {font=\relsize{#1}}}
\usetikzlibrary{calc}
\graphicspath{{img/}}

\begin{document}
\title{Controllable Capillary Assembly of Magnetic 
Ellipsoidal Janus \\ Particles into Tunable Rings, Chains and Hexagonal Lattices}
 \author{Qingguang Xie}
  \affiliation{Department of Applied Physics, Eindhoven University of Technology, P.O. Box 513, 5600MB Eindhoven, The Netherlands}
  \author{Jens Harting}
  \email{j.harting@fz-juelich.de}
  \affiliation{Helmholtz Institute Erlangen-N\"urnberg for Renewable Energy (IEK-11), Forschungszentrum J\"ulich, F\"urther Str. 248, 90429 N\"urnberg, Germany}
\affiliation{Department of Chemical and Biological Engineering and Department of Physics, Friedrich-Alexander-Universit\"at Erlangen-N\"urnberg, F\"urther Str. 248, 90429 N\"urnberg, Germany}
\begin{abstract}
Colloidal assembly at fluid interfaces has a great potential for the bottom-up
	fabrication of novel structured materials.  However, challenges remain
	in realizing controllable and tunable assembly of particles into
	diverse structures. Herein, we report the capillary assembly of
	magnetic ellipsoidal Janus particles at a fluid-fluid interface.
	Depending on their tilt angle, i.e. the angle the particle main axis
	forms with the fluid interface, these particles deform the interface
	and generate capillary dipoles or hexapoles. Driven by capillary
	interactions, multiple particles thus assemble into chain-, hexagonal
	lattice- and ring-like structures, which can be actively controlled by
	applying an external magnetic field. We predict a field-strength phase
	diagram in which various structures are present as stable states. Owing
	to the diversity, controllability, and  tunability of assembled
	structures, magnetic ellipsoidal Janus particles at fluid interfaces
	could therefore serve as versatile building blocks for novel materials.
\end{abstract}

\maketitle
\input{intro.tex}
\input{single-particle.tex}

\input{more-particles.tex}

\input{final.tex}

\vspace*{2mm}
\noindent
\textbf{\large Supporting Information} \par 
\noindent
Supporting Information is available.
\vspace*{2mm}\\
\noindent
\textbf{\large Acknowledgements}\\
We thank Oscar Coppelmans for fruitful discussions.
Financial support is acknowledged from the Netherlands Organization for
	Scientific Research (NWO) through a NWO Industrial Partnership
	Programme (IPP).  This research programme is co-financed by
	Canon Production Printing Netherlands B.V., University of Twente and Eindhoven University of
	Technology. We are also grateful for financial support from the German
	Research Foundation (DFG) through priority program SPP2171 (grant HA
	4382/11-1).  We thank the J\"ulich Supercomputing Centre and the High
	Performance Computing Center Stuttgart for the technical support and
	allocated CPU time. 
\vspace*{2mm}\\
\textbf{\large Conflict of Interest}\\
The authors declare no conflict of interest. \\
\vspace*{2mm}

\end{document}

%% file: intro.tex
Self-assembly of particles has received considerable attention with respect to
their potential applications in advanced technologies, such as displays,
sensors and optoelectronic devices. For certain industrial applications
including large-area coatings, the assembly of particles can prove far more
efficient than “top-down” approaches such as lithography. However, controlling
particles into highly tunable and predictable structures remains a
challenge.\citeref{Weiss2008,Grzelczak2010,Park2014review,Toor2016} Herein, we
report a highly controllable and versatile strategy for the fabrication of
ordered structures via capillary assembly of magnetic ellipsoidal Janus
particles at a fluid-fluid interface.

Colloidal particles strongly attach at a fluid-fluid interface\citeref{Binks2001}
and deform the interface due to their weight, anisotropic shape and
roughness.\citeref{Chan1981,Lucassen1992,Stamou2000,Danov2005}
If neighbouring particles generate deformations of the interface that overlap,
capillary interactions arise which drive the particles to assemble into ordered
structures.  Such capillary interactions can be characterized by the modes of
interface deformation.  In the limit of small interface deformation, the
interface height $\mathbf{h}$ around the particle can be described according to
the Young-Laplace equation $\nabla \mathbf{h} = 0$ and it can be analysed by a
multipole expansion in analogy with
electrostatics\citeref{Lehle2008,Stamou2000}.  In the case of heavy
particles attached to the interface, the interface is pushed down by the
particles resulting in a monopolar deformation\citeref{Chan1981}. For
particularly small or light particles, the contact line where particle and
fluid-fluid interface meet can undulate due to anisotropic shapes or the
roughness of the particle surface and, thus, induce quadrupolar or hexapolar
interface deformations\citeref{Lucassen1992,Stamou2000,Danov2005}. Driven by the
capillary interactions, particles assemble into specific structures to reduce
the total adsorption free energy.  For example, heavy particles tend to form a
cluster\citeref{Bowden1997}, while ellipsoidal particles assemble into
chains\citeref{Loudet2005,Loudet2006a}. Furthermore, cubic particles generate
hexagonal or honeycomb lattices\citeref{Rene2016}. However, such structures are
not dynamically tunable because the capillary interactions are dependent on the
intrinsic properties of the particles, such as their weight and their precise
shape.

The synthesis of colloidal particles with specific physical properties (e.g.,
electric or magnetic moments) and anisotropic chemical properties (e.g.,
amphiphilic Janus particles) interacting with external fields allows for a
greater control of the assembly process. For example, magnetic
ellipsoids\citeref{Gary2014a,Gary2014b} and magnetic spherical Janus
particles\citeref{Xie2015} at fluid interfaces can generate dipolar capillary
interactions. Those interactions can be precisely controlled by an external
magnetic field. Yet, to the best of our knowledge, so far the assembly of
particles was limited to create specific and regular structures, such as
chains\citeref{Gary2014b,Xie2016} or hexagonal lattices\citeref{Xie2017ACSNano}. 

Here, we examine a combination of anisotropic physical properties and chemical
properties of particles to achieve the formation of diverse structures with
controllability and tunablity.  We investigate the behavior of magnetic
ellipsoidal Janus particles at a flat fluid-fluid interface, interacting with
external magnetic fields.  We find that by varying its tilt angle due to the
presence of an external magnetic field, a single particle generates a dipolar
or a hexapolar interface deformation.  Driven by the resulting dipolar or
hexapolar capillary interactions, such particles assemble into rings, chains,
and hexagonal lattice structures, where a dynamical transition between these
assembly states can be obtained by dynamically manipulating the field.

%% file: single-particle.tex
\begin{figure*}[]
\begin{subfigure}{.23\textwidth}
  \includegraphics[height = 0.99\textwidth]{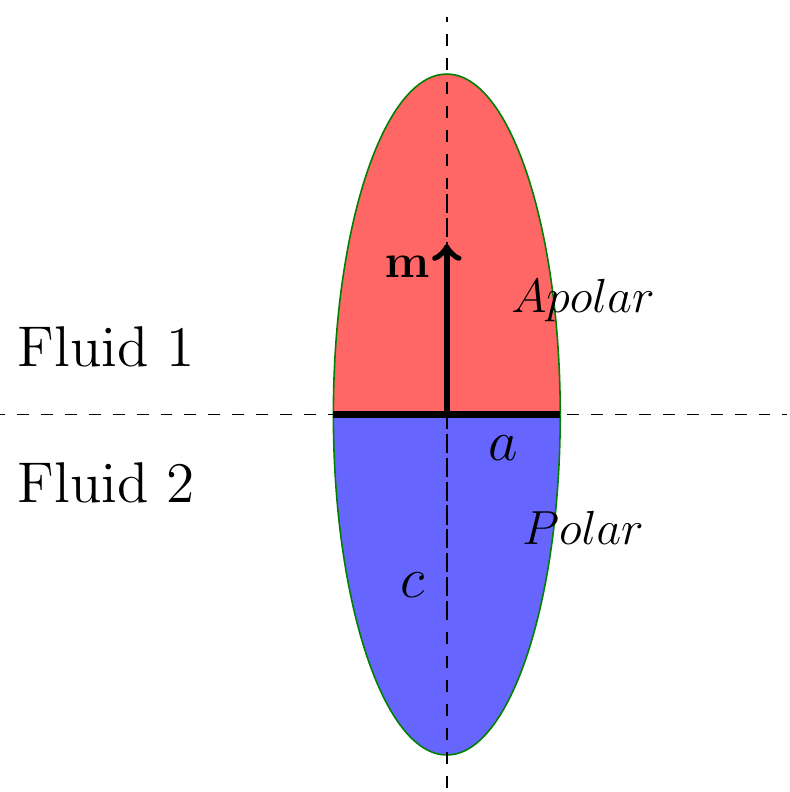}
  \caption{Upright orientation}
 \label{fig:sketch-up}
\end{subfigure}
\begin{subfigure}{.23\textwidth}
  \includegraphics[height = 0.99\textwidth]{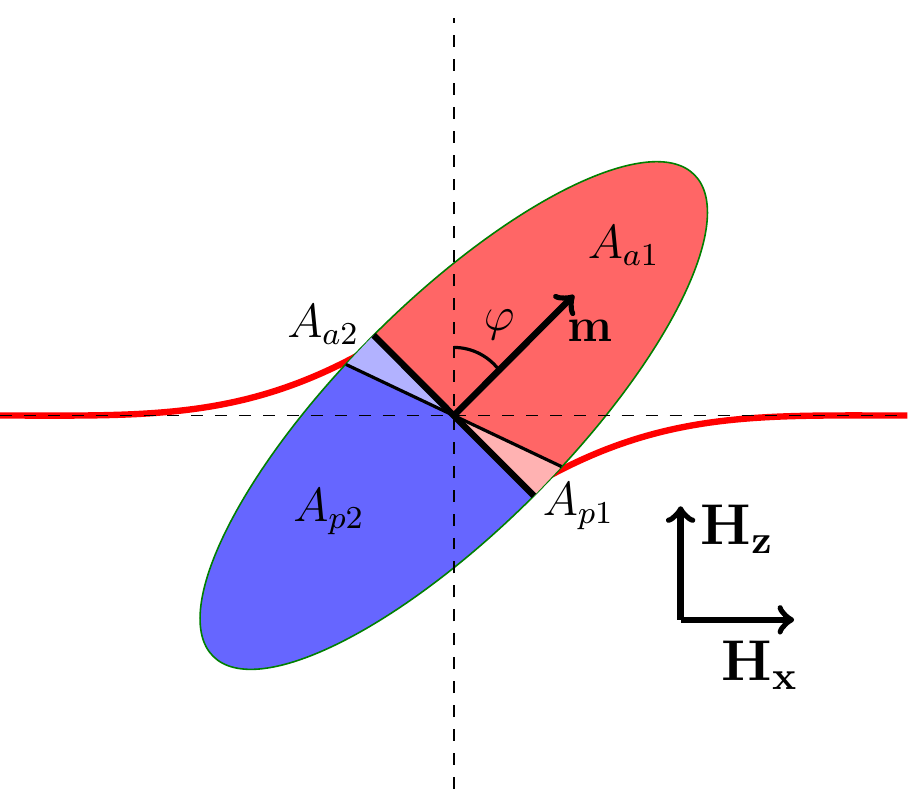}\\
  \caption{Tilted orientation}
 \label{fig:sketch}
\end{subfigure}
\hspace{4mm}
\begin{subfigure}{.23\textwidth}
  \includegraphics[width = 0.99\textwidth]{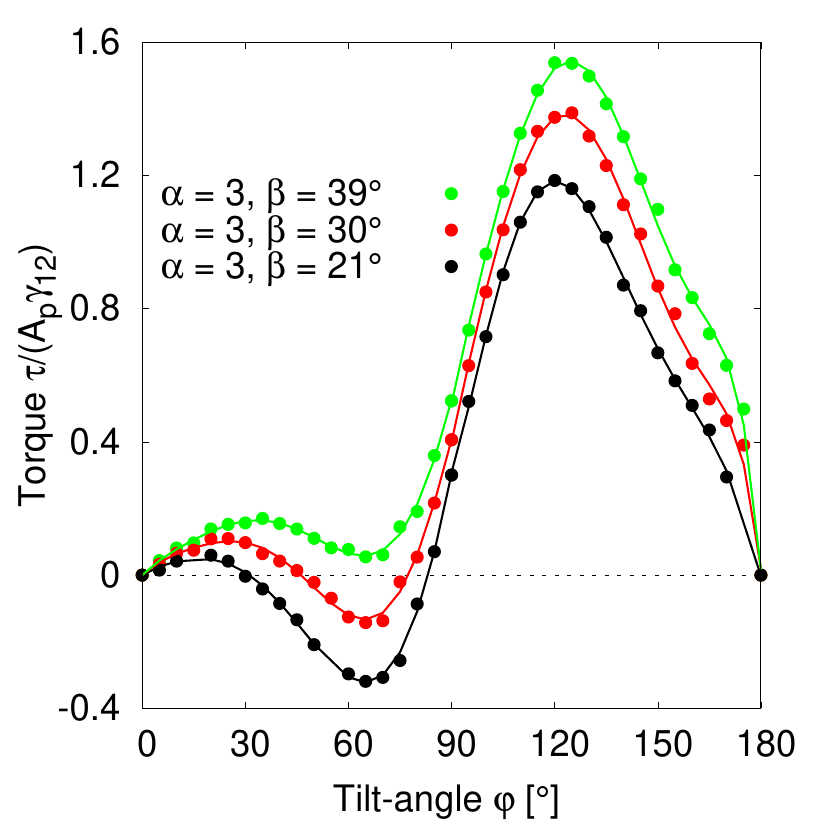}
  \caption{Torque}
 \label{fig:angle_tor}
\end{subfigure}
\begin{subfigure}{.23\textwidth}
  \includegraphics[width = 0.99\textwidth]{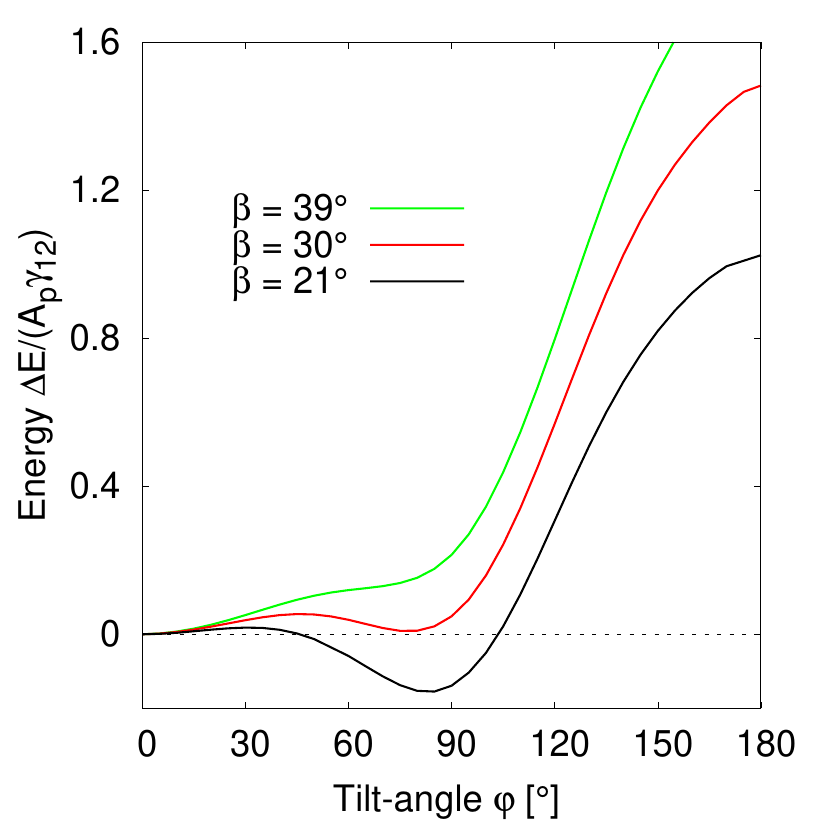}
  \caption{Energy}
 \label{fig:energy_tor}
\end{subfigure}
\caption{A single ellipsoidal Janus particle adsorbed at a fluid-fluid interface in an upright orientation a) and in a tilted orientation b). 
c) Reduced torque $\tau/A_p\gamma_{12}$ and d) free energy $\Delta E/A_p\gamma_{12}$ of a Janus ellipsoid with aspect ratio $\alpha=3$ 
as a function of tilt angle $\varphi$  for  $\beta = 21^{\circ}$ (black) , 
$\beta = 30^{\circ}$ (red)  and $ \beta = 39^{\circ}$ (green), where $A_p = \pi ac$ and $\gamma_{12}$ is the fluid-fluid interface tension. 
}
\label{fig:geo}
\end{figure*}
We consider an ellipsoidal Janus particle adsorbed at a fluid-fluid interface,
as illustrated in \figref{sketch-up} and \figref{sketch}.  The particle is
composed of apolar and polar hemispheres of opposite wettability, represented
by the three-phase contact angles $\theta_A = 90^{\circ} +\beta$  and $\theta_P
= 90^{\circ} -\beta$, respectively, where $\beta$ indicates the amphiphilicity
of the particle. The boundary between these two hemispheres is called the Janus
boundary.  We denote the radii of long- and short-axes of the Janus ellipsoid
with $c$ and $a$, respectively, and the aspect ratio $\alpha$ is defined as
$\alpha=c/a$.  The magnetic moment $m$ is perpendicular to the Janus boundary,
and external magnetic fields $\mathbf{H}_x$ and $\mathbf{H}_z$ are applied in
horizontal and vertical direction, respectively. We define magnetic
dipole-field strengths
$B_x=|\mathbf{H_x}||\mathbf{m}|$ and $B_z=|\mathbf{H_z}||\mathbf{m}|$, which
represent the magnitude of the interactions between the magnetic dipole and the
external fields. We apply lattice Boltzmann
simulations\citeref{Jansen2011,Gunther2013,Xie2015} to
investigate the behaviour of magnetic ellipsoidal Janus particles adsorbed at a
liquid-liquid interface.  A detailed description of the method and simulation
parameters is provided in the Supplementary Material.
In our simulations, magnetic dipole-dipole interaction forces are 
six orders of magnitude smaller than the capillary interaction forces.
Therefore, the dipole-dipole interaction is effectively negligible, and the
assembling of structures is purely dominated by the capillary interactions
between particles. 

In the absence of an external magnetic field, an isolated ellipsoidal Janus
particle adsorbed at an interface takes its equilibrium orientation to minimize
the total adsorption free energy\citeref{Park2012a,Rezvantalab2013a}. 
The total adsorption free energy is written as $E = \gamma_{12}A_{12} +
\gamma_{a1}A_{a1} + \gamma_{p2}A_{p2}  +\gamma_{a2}A_{a2} + \gamma_{p1}A_{p1}
$, where $\gamma_{ij}$ are the interface tensions between phases $i$ and $j$
and $A_{ij}$ are the contact surface areas between phases $i$ and $j$, where
$i,j$ $=$ $\{1$: fluid, $2$: fluid, $a$: apolar, $p$: polar$\}$. 
There is no exact analytical expression for the free energy of a tilted Janus
ellipsoid at an interface, due to the difficulty in modelling the shape of the
deformed interface and the segment area of the ellipsoid.  Our lattice
Boltzmann simulations are capable of capturing interface deformations fully
without imposing any assumptions about the magnitude of the deformations or
stipulating any particle-fluid boundary conditions\citeref{Xie2015}. 
We take the upright orientation $\theta=0$ as a reference configuration, and
numerically calculate the free energy. 
To do this, we initialize the particle on the interface with the desired tilt
angle and then fix the position and orientation of it. After equlibration of
the fluids, we measure the torque subjected on the particles from the
fluid-fluid interface in the absence of magnetic fields. We then obtain the
free energy $\Delta E = E_{\varphi}-E_{\varphi=0}$ by integrating the torque on
the particle as the particle rotates quasi-statically, $\Delta E =
\int_{0}^{\varphi_{\text{tilt}}} \tau_{\varphi} d\varphi$.

\begin{figure*}[t!]
 \begin{subfigure}{.3\textwidth}
\includegraphics[width= 0.95\textwidth]{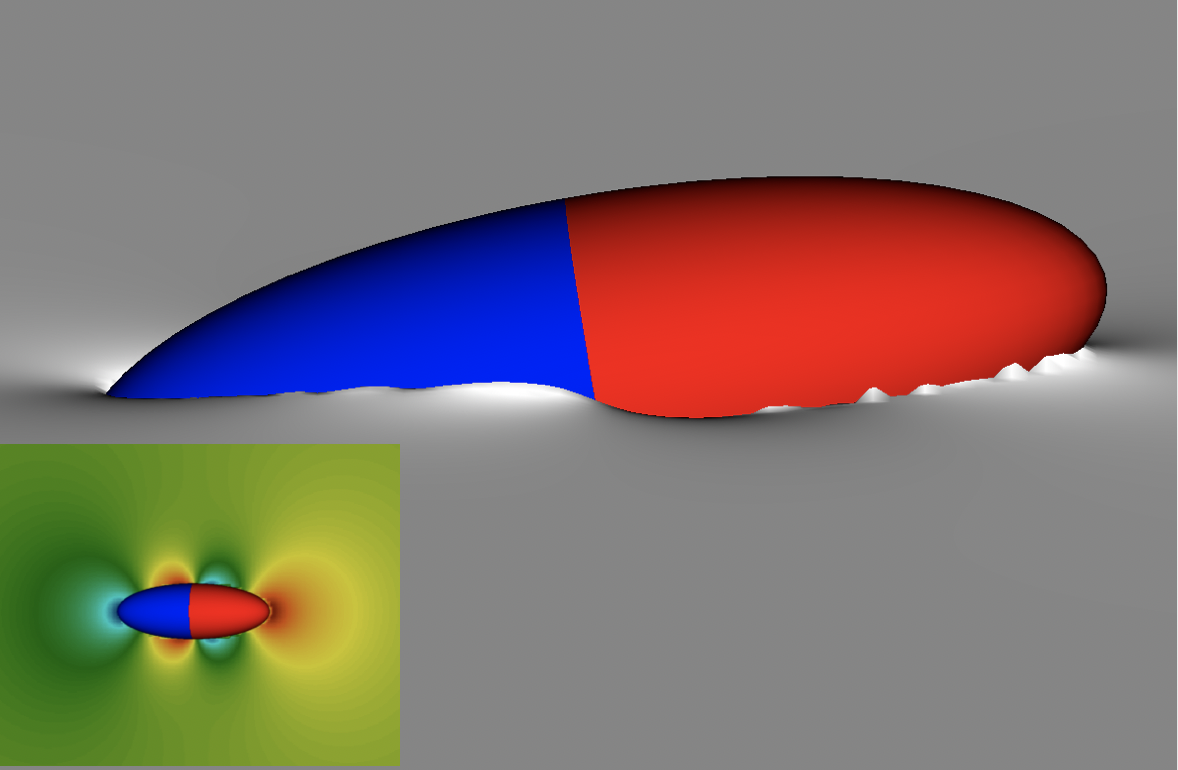}
 \subcaption{$\varphi=80^{\circ}$}
 \label{fig:def-1}
 \end{subfigure}
 \begin{subfigure}{.3\textwidth}
\includegraphics[width= 0.95\textwidth]{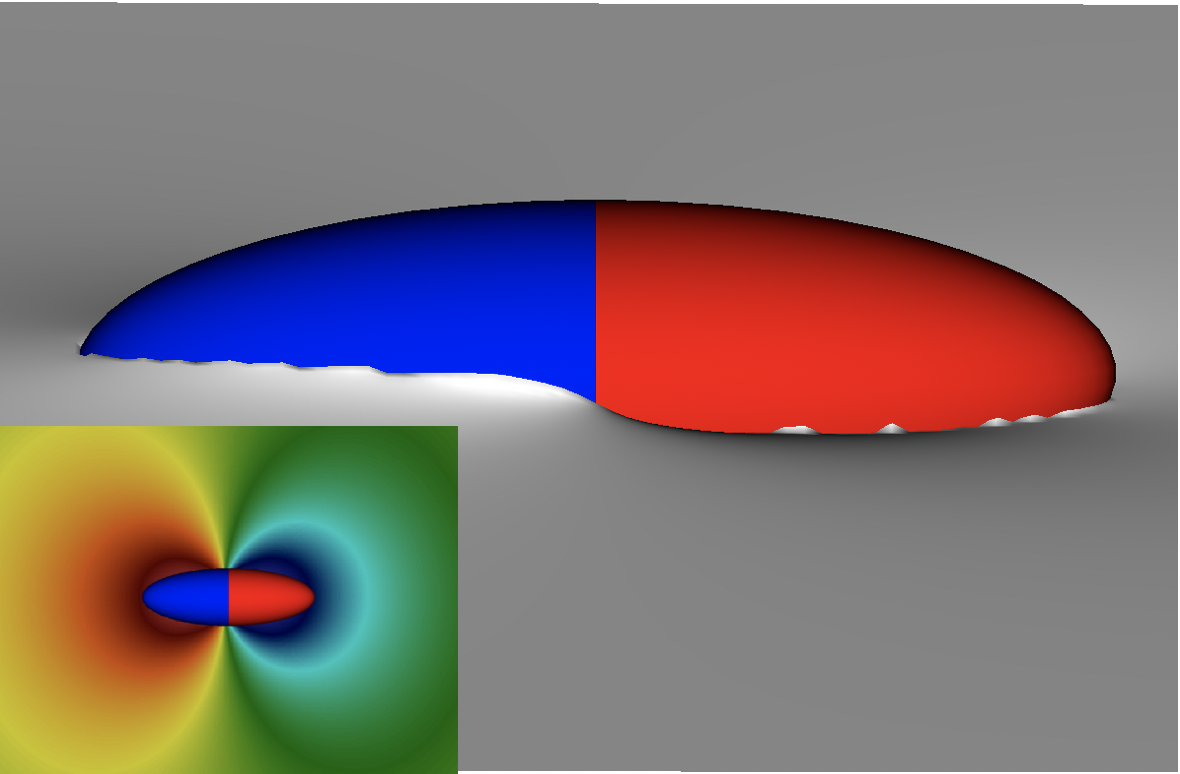}
 \subcaption{$\varphi=90^{\circ}$}
 \label{fig:def-2}
 \end{subfigure} 
 \begin{subfigure}{.3\textwidth}
\includegraphics[width= 0.95\textwidth]{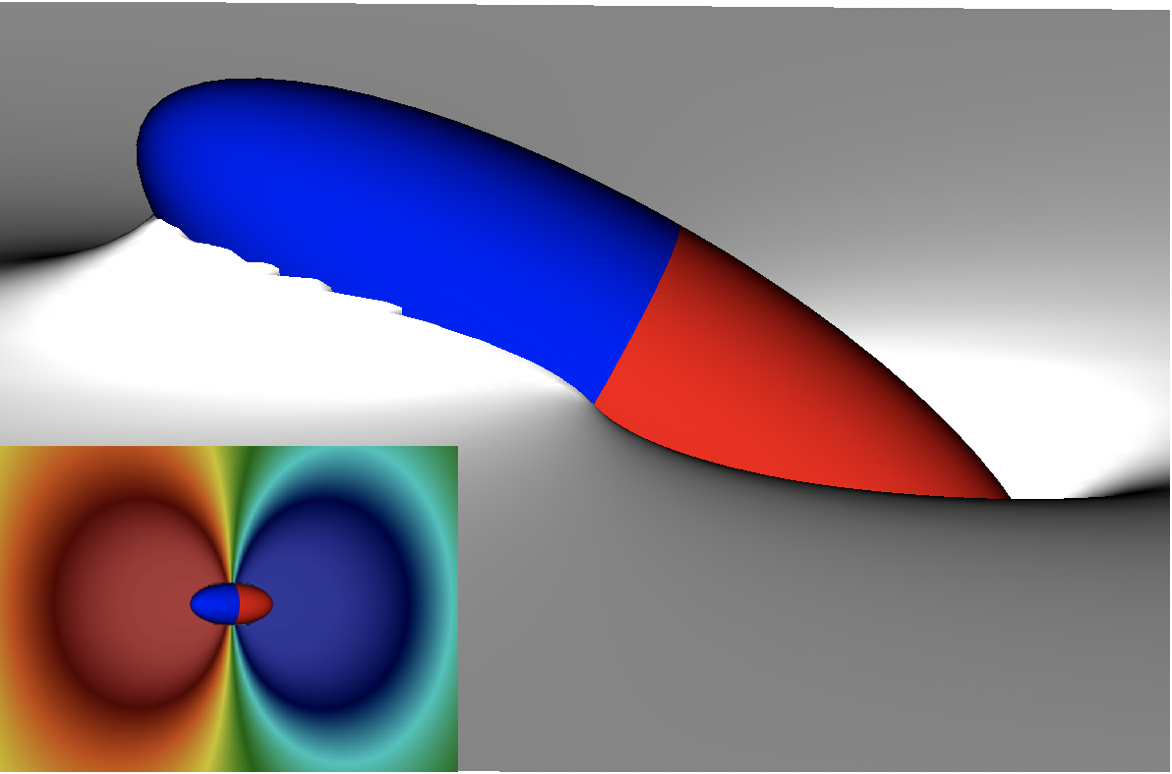}
 \subcaption{$\varphi=120^{\circ}$}
 \label{fig:def-3}
 \end{subfigure}
 
 \caption{Snapshots of an ellipsoidal Janus particle at a fluid-fluid interface
	at different tilt angles as obtained from our simulations. The
	interface deformation appears to be hexapolar at $\varphi=80^{\circ}$
	(a), dipolar at $\varphi=90^{\circ}$ (b), and also dipolar at
	$\varphi=120^{\circ}$, but with a larger deformation (c). The insets
	depict the hexapolar/dipolar interface deformation.
	}
\label{fig:def}
 \end{figure*} 

\figref{angle_tor} shows the evolution of this torque $\tau_{\varphi}$ versus
the tilt angle of a Janus ellipsoid (aspect ratio $\alpha=3$) for different
amphiphilicities $\beta = 21^{\circ}$ (black), $\beta = 30^{\circ}$ (red) and $
\beta = 39^{\circ}$ (green). For all amphiphilicities, the torque is zero at
$\theta=0^{\circ}$ and increases linearly for small tilt angles ($\varphi <
30^{\circ}$), in the direction resisting the rotation of the particle. As the
tilt angle increases further $\varphi \to 70^{\circ}$, the torque decreases
followed by a sharp increase until the tilt angle $\varphi \to 130^{\circ}$.
Finally, the torque decreases to zero when the tilt angle approaches
$\varphi=180^{\circ}$.

\figref{energy_tor} shows the free energy $\Delta E$ of the ellipsoidal Janus
particle, as a function of the tilt angle for the same amphiphilicities as in
\figref{angle_tor}.  For a large amphiphilicity $\beta=39^{\circ}$ the free
energy keeps increasing for the whole range of tilt angles, indicating that the
particle in the upright orientation $\varphi=0^{\circ}$ corresponds to the
global energy minimum. The particle tends to reduce the free energy by
increasing the interfacial area between the particle and its preferred fluid
phase.  For smaller amphiphilicities $\beta=21^{\circ}$ and $\beta=30^{\circ}$,
the free energy is not monotonic: for small tilt angles $\theta<30^{\circ}$,
the free energy increases, followed by a decrease until the tilt angle
increases further to $\theta \sim 80^{\circ}$,  and afterwards, the free energy
continuously increases until the tilt angle reaches $\theta=180^{\circ}$.  For
an amphiphility $\beta=21^{\circ}$, \figref{energy_tor} indicates the presence
of a local energy minimum for particles in the upright orientation
($\varphi=0^{\circ}$) and a global energy minimum for particles in the tilted
state ($\varphi\sim 80^{\circ}$). The Janus ellipsoid tends to occupy more of
the fluid-fluid surface area and in turn reduces the free energy.  An energy
barrier exists between the metastable upright orientated configuration and the
global energy minimum, which requires a magnetic torque $\tau_{m}$ stronger
than the capillary torque $\tau/A_p \gamma \sim 0.35$ (as shown in
\figref{angle_tor}), to rotate the particle out of the tilted equilibrium
orientation to the upright orientation.

\begin{figure*}[ht]
  \begin{subfigure}{.19\textwidth}
\includegraphics[width= 0.98\textwidth]{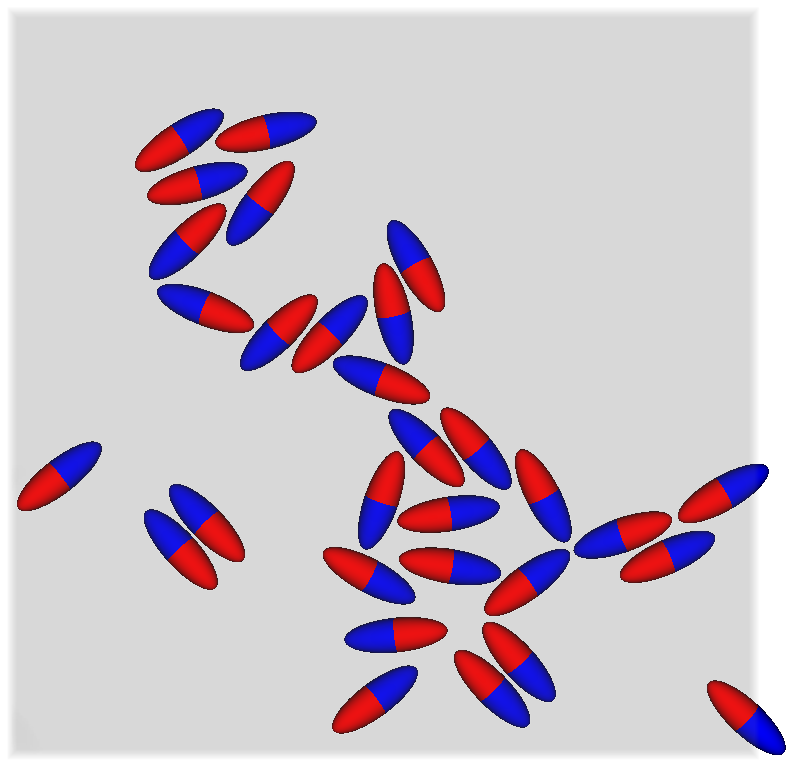}
 \subcaption{$\bar{B}_x=\bar{B}_z=0$}
 \label{fig:p30-1}
 \end{subfigure}
  \begin{subfigure}{.19\textwidth}
\includegraphics[width= 0.98\textwidth]{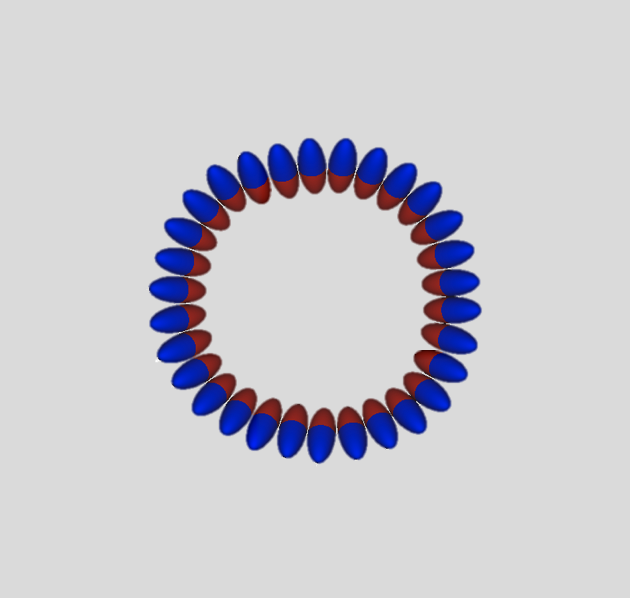}
 \subcaption{$\bar{B}_x=0,\bar{B}_z=-1.3$}
 \label{fig:p30-2}
 \end{subfigure}
   \begin{subfigure}{.19\textwidth}
\includegraphics[width= 0.98\textwidth]{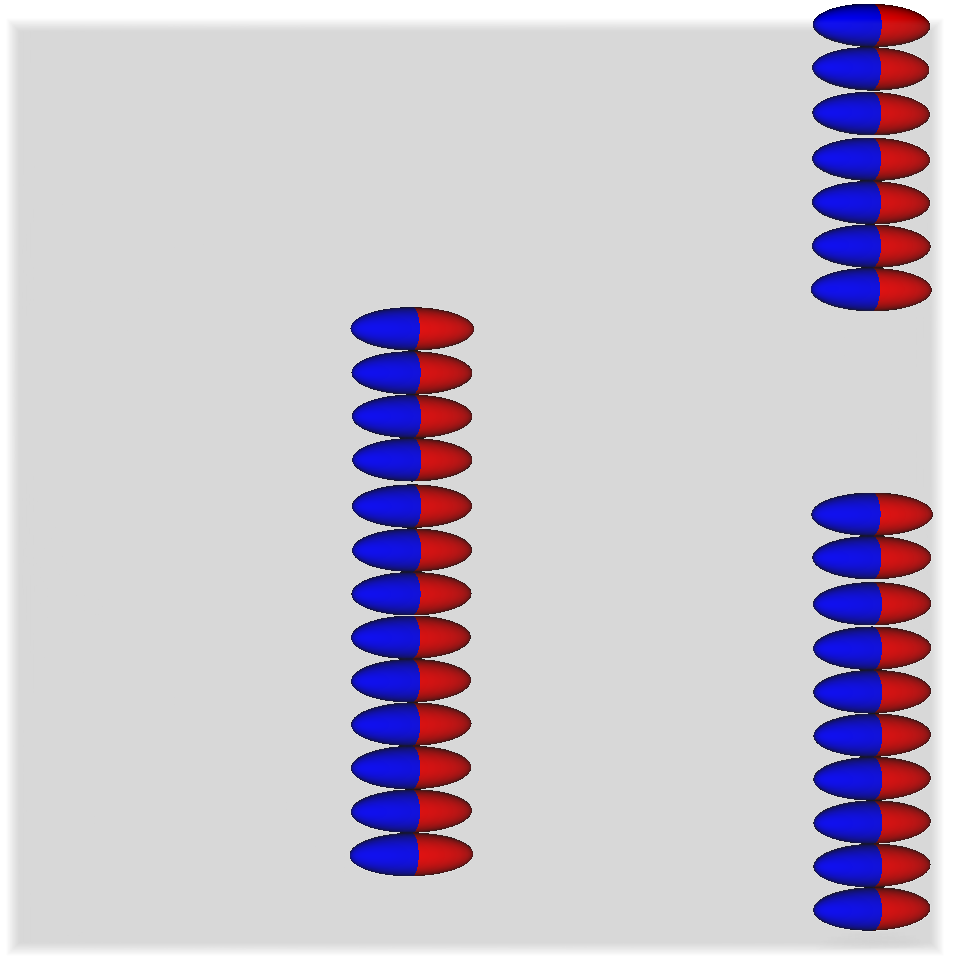}
 \subcaption{$\bar{B}_x=-\bar{B}_z=1.3$ }
 \label{fig:p30-3}
 \end{subfigure}
 \begin{subfigure}{.19\textwidth}
\includegraphics[width= 0.98\textwidth]{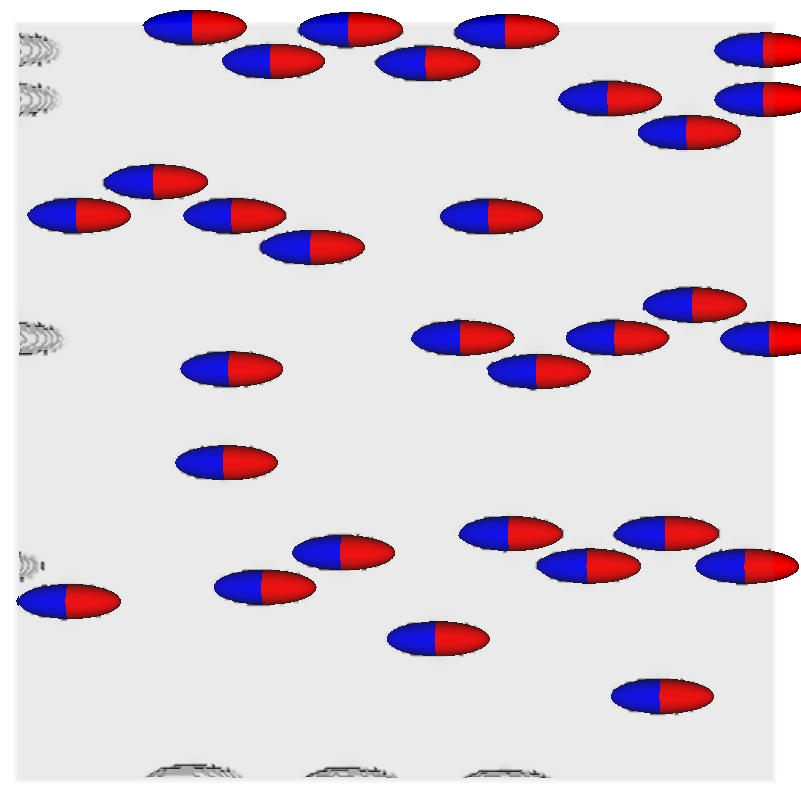}
 \subcaption{$\bar{B}_x=1.3,\bar{B}_z=0.18$ }
 \label{fig:p30-4}
 \end{subfigure}
  \begin{subfigure}{.19\textwidth}
\includegraphics[width= 0.98\textwidth]{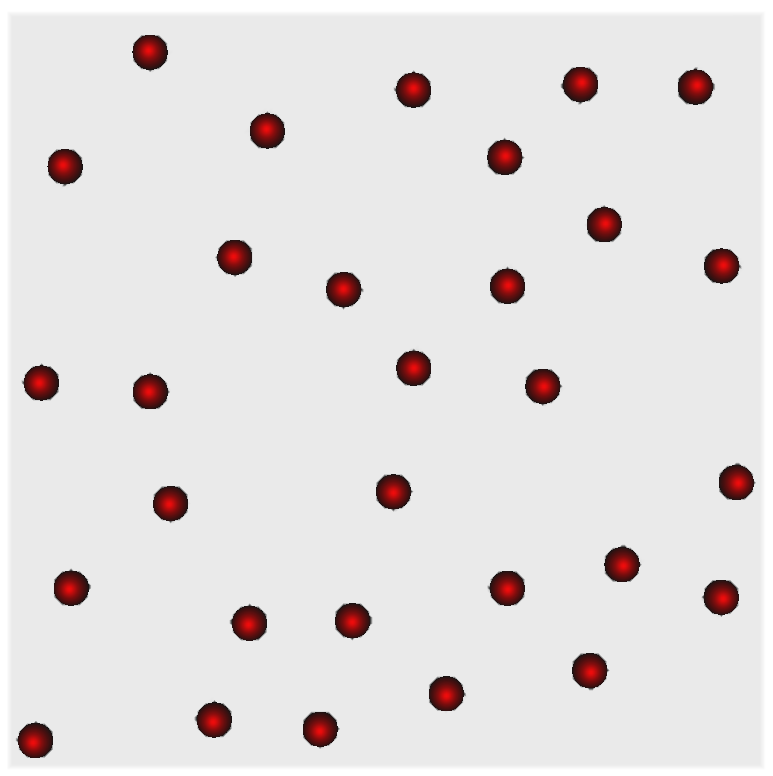}
 \subcaption{$\bar{B}_x=0,\bar{B}_z=0.6$ }
 \label{fig:p30-5}
 \end{subfigure}
 
  \begin{subfigure}{.19\textwidth}
\includegraphics[width= 0.98\textwidth]{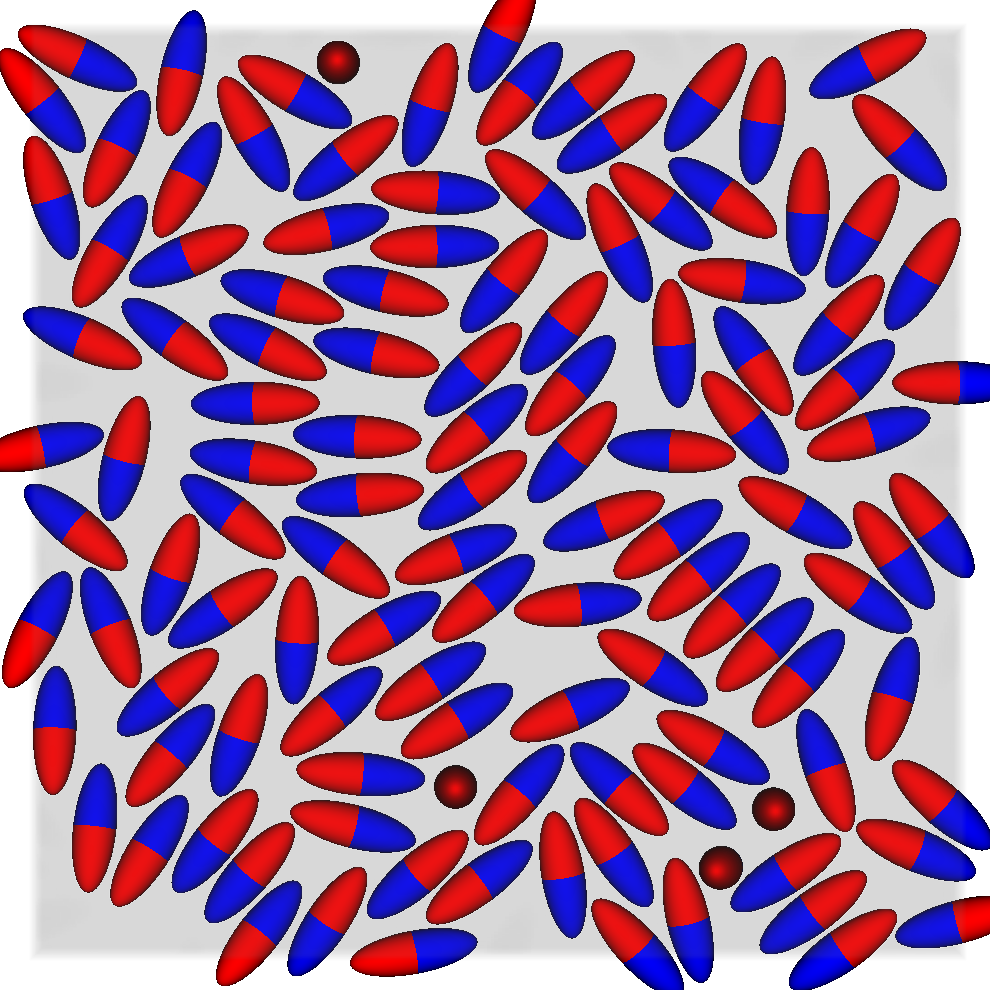}
 \subcaption{$\bar{B}_x=0,\bar{B}_z=0$}
 \label{fig:p120-1}
 \end{subfigure}
 \begin{subfigure}{.19\textwidth}
\includegraphics[width= 0.98\textwidth]{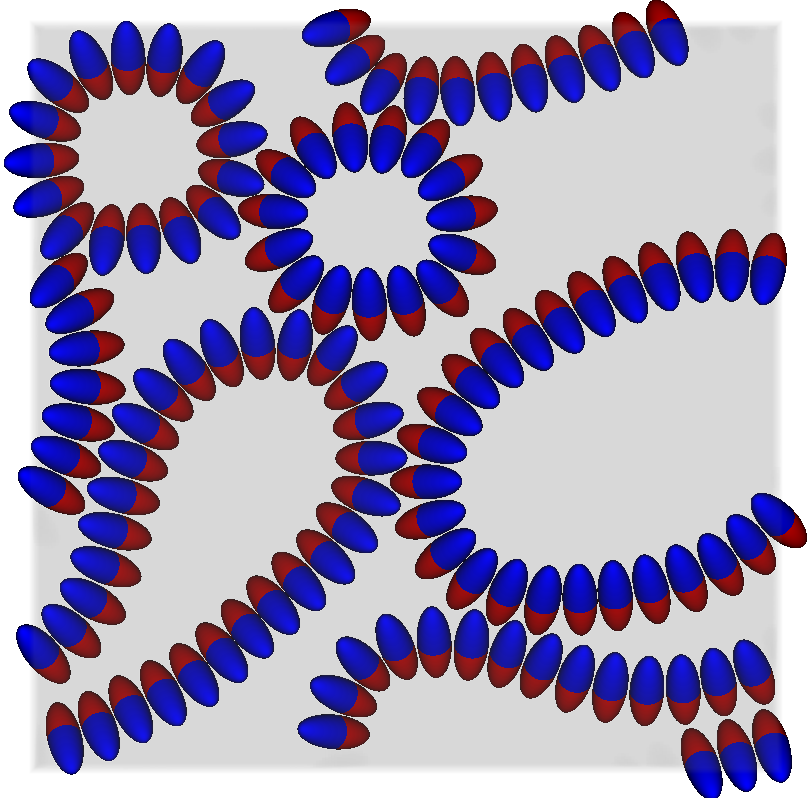}
 \subcaption{$\bar{B}_x=0,\bar{B}_z=-1.3$ }
 \label{fig:p120-2}
 \end{subfigure}
   \begin{subfigure}{.19\textwidth}
\includegraphics[width= 0.98\textwidth]{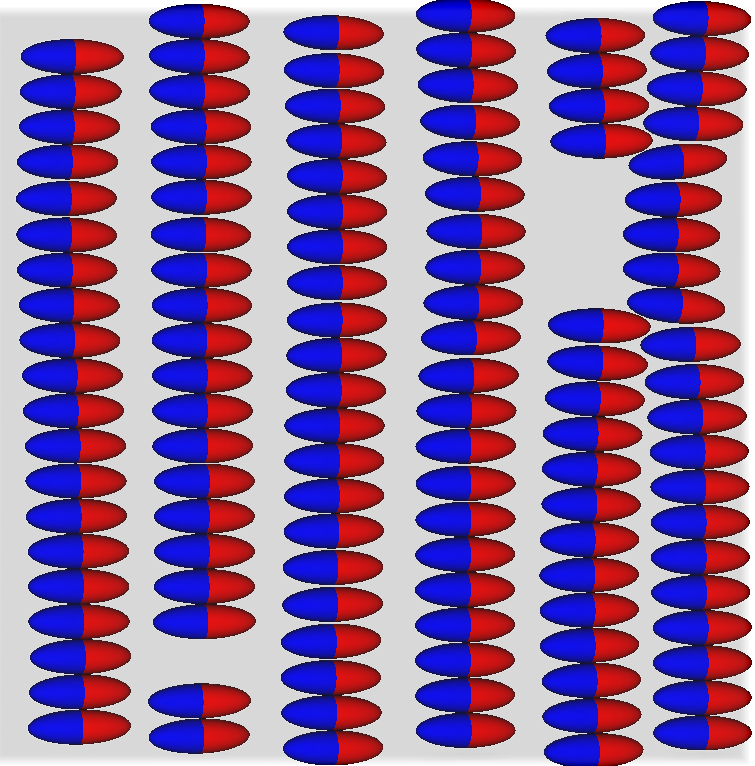}
 \subcaption{$\bar{B}_x=-\bar{B}_z=1.3$}
 \label{fig:p120-3}
 \end{subfigure}
  \begin{subfigure}{.19\textwidth}
\includegraphics[width= 0.98\textwidth]{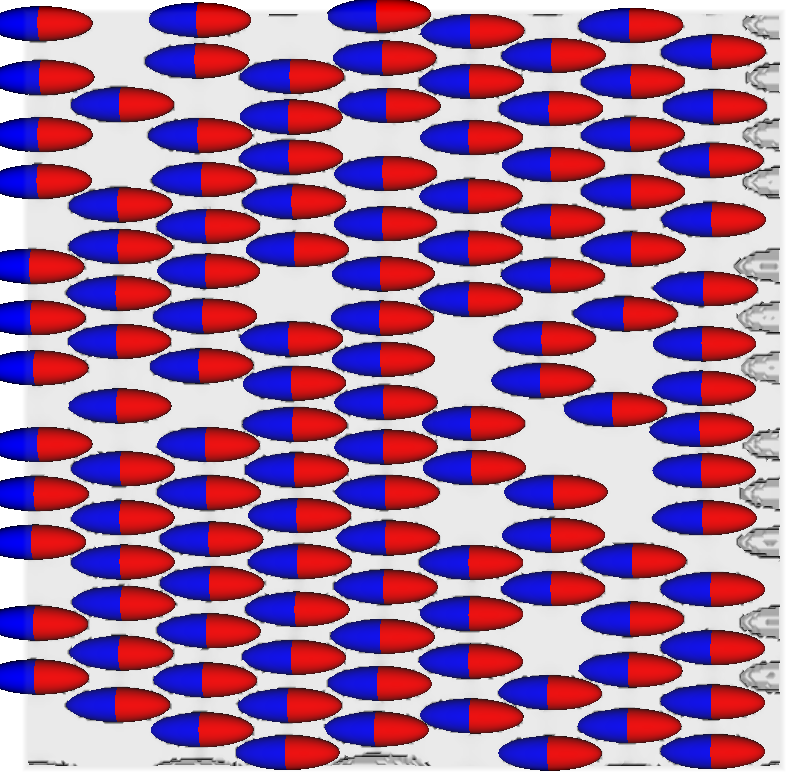}
 \subcaption{$\bar{B}_x=1.3,\bar{B}_z=0.18$ }
 \label{fig:p120-4}
 \end{subfigure}
   \begin{subfigure}{.19\textwidth}
\includegraphics[width= 0.98\textwidth]{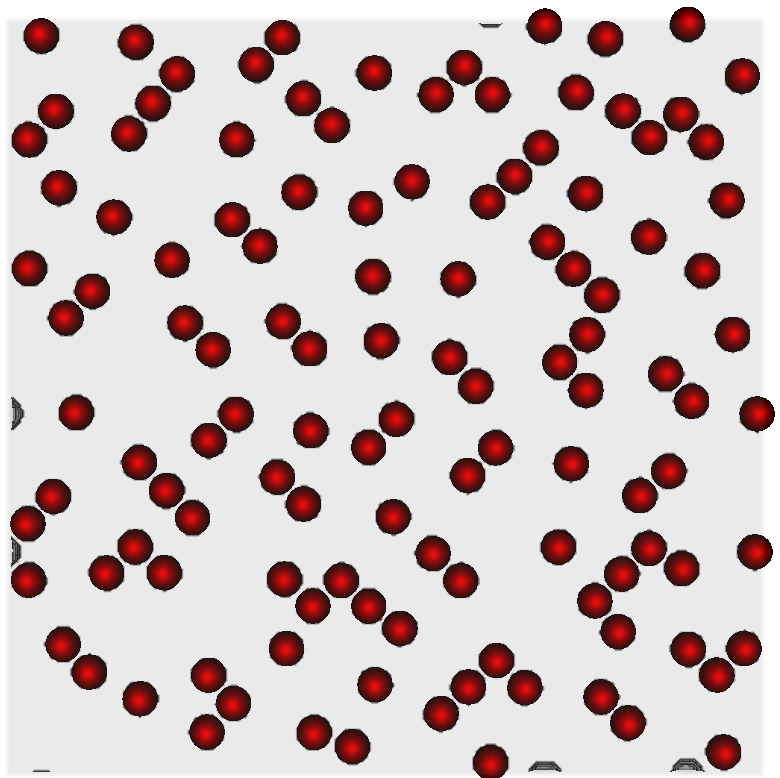}
 \subcaption{$\bar{B}_x=0,\bar{B}_z=0.6$ }
 \label{fig:p120-5}
 \end{subfigure}

 \caption{Assembly of particles with different surface fractionr, a-e)
 $\Phi=0.16$ and f-j) $\Phi=0.62$ under different magnetic fields. The particles
 have an aspect ratio $\alpha=3$ and an amphiphilicity $\beta=21^{\circ}$. 
 }
\label{fig:p30-120}
 \end{figure*}

Next, we investigate the interface deformation induced by the ellipsoidal Janus
particle at different tilt angles.  
We find that the interface stays undeformed around the particle with upright
orientation, indicating the absence of a
torque at $\varphi=0^{\circ}$, which is consistent with our results
in~\figref{angle_tor}.  \figref{def} shows how the three-phase contact line and
interface deform around the ellipsoidal Janus particle ($\alpha=3,
\beta=21^{\circ}$) for different tilt angles $\varphi=80^{\circ}$,
$\varphi=90^{\circ}$ and $\varphi=120^{\circ}$, respectively.  At
$\varphi=80^{\circ}$, the interface deforms around the particle in a hexapolar
shape (\figref{def-1}), with three rises and three dips distributed around the
particle.  The interface is raised up at the tip of the apolar hemisphere and
depressed at the tip of the polar hemisphere.  When the particle aligns in the
horizontal orientation $\varphi=90^{\circ}$, the interface shows a dipolar
deformation (\figref{def-2}), with a dip around the apolar hemisphere and a
rise around the polar hemisphere. We note that the magnitude of this dipolar
deformation is much stronger than that of a hexapolar deformation. 
\revisedtext{Furthermore, the dependence of the mode of deformation on the particle orientation 
cannot be generated by Janus spheres~\cite{Xie2015} or homogeneous ellipsoids~\cite{Gary2014a}. 
This dependence was also not discovered by previous works on Janus ellipsoids~\cite{Park2012a,Rezvantalab2013a}, where only the equilibrium orientation of Janus ellipsoids were discussed.}
The deformation increases further with increasing tilt angle to
$\varphi=120^{\circ}$ and the deformed interface height is even at the same
order of particle short radius (\figref{def-3}).  We also observe an
unsymmetrical hexoplar, butterfly-like deformation at intermediate tilt angles
(Supplmentary Material Figure S1).  The rise and dip generated around the Janus ellipsoid result from
the competition between Janus and ellipsoidal properties of the particle: it is
known that both, a Janus sphere and a homogeneous ellipsoid, generate dipolar
interface deformations. However, the rise and dip areas are located
oppositely\citeref{Xie2015,Gary2014a}.  The positioning and strength of the rises
and dips can be tuned by varying the aspect ratio and amphiphilicity of the
particles, as observed in our simulations (Figure S2). \revisedtext{With increasing either the aspect ratio or the amphiphilicity, the dipolar deformation
is expected to dominate for the full range of tilt angles and the hexagonal deformation becomes negligible.}

%% file: more-particles.tex
If two or more particles are adsorbed at a fluid-fluid interface, the
deformations induced by neighbouring particles can overlap, leading to
capillary interactions between these particles. A pair of Janus ellipsoids
interacting through hexapolar or dipolar capillary interactions prefers to
align in a side-side configuration (Figure~S3), corresponding to a capillary
energy minimum configuration\citeref{Rezvantalab2013a, Xie2015}. However,
many-body effects in the capillary assembling of multiple Janus ellipsoids
under external magnetic fields remain to be explored. In~\figref{p30-120} we
show the assembled structures for particle surface fractions $\Phi = 0.16,
0.62$ that form as we vary the dipole-field strengths $B_x$ and $B_z$. We
define the particle surface fraction as $\Phi = N \pi ac/A$, where $N$ is the
total number of particles and $A$ is the interface area before particles are
placed at the interface. The particles have an aspect ratio $\alpha=3$ and an
amphiphilicity $\beta=21^{\circ}$.  We define the normalized dipole-field
strength as $\bar{B}_i=B_i/A_p\gamma_{12}$, where $i=x,z$. Initially, the
particles are distributed randomly on the interface.  In the absence of
external magnetic fields, they take their tilted equilibrium orientation
$\varphi \sim 80^{\circ}$ and introduce hexapolar interface deformations (as
shown in~\figref{def-1}). Then, the particles assemble into locally-ordered
structures (\figref{p30-1} and \figref{p120-1}).
Small particle clusters with side-side, tip-tip, and side-tip alignments
coexist, which indicates that multiple particles interacting through hexapolar
capillary interactions have different (meta)stable configurations.  
When applying a downward magnetic field $\bar{B}_z=-1.3$, the particles tilt at
$\varphi \sim 140^{\circ}$ and generate dipolar interface deformations (as
shown in \figref{def-3}).
For a lower particle surface fraction $\Phi=0.16$, the particles form a
circular ring (\figref{p30-2}), instead of a chain structure predicted by the
pair-wise interactions, indicating the presence of strong many-body
interactions. Our results are consistent with the theoretical prediction that a
closed loop structure is the capillary energy minimum configuration for
multiple tilted ellipsoidal particles interacting with dipolar capillary
interactions\citeref{Buzza2018}.  With increasing the surface fraction
$\Phi=0.62$, the particles form multiple rings and the rings are more curved
due to geometrical restriction. Along with $\bar{B}_z=-1.3$, additionally we
apply a horizontal magnetic field $\bar{B}_x=1.3$. Then, the particles align
into chain-like structures for both lower and higher surface fractions
(\figref{p30-3} and \figref{p120-3}), consistent with the prediction from
pair-wise interactions, demonstrating that many-body effects are less relevant
in this case. \revisedtext{We note that both attractive and repulsive capillary forces are present between 
neighboring particles dependent on their relative orientations. 
The particles in the same chain touch each other due to attractive forces, whereas, 
the neighboring chains repel each other, resulting in a clear 
separation of chains and easy rearrangement of particles with varying external fields.}
Here, the particles are forced to align in the direction parallel
to the horizontal magnetic field and they only have $2$ degrees of freedom
(translation in $x$ and $y$ directions), which weakens the many-body effect. On
the contrary, when only a vertical magnetic field $\bar{B}_z$ is applied, the
particles have $3$ degrees of freedom (translation in $x$, $y$ direction and
rotation around the $z$ axis) and capillary torques can rotate the particles to
form rings. If an upward magnetic field $\bar{B}_z=0.18$ together with a
horizontal field $\bar{B}_x=1.3$ is applied, the particles show a tilt angle
$\varphi  \sim 80^{\circ}$ and generate hexapolar deformations. For lower
particle surface fraction $\Phi=0.16$, the particles align in a zigzag
structure (\figref{p30-4}).
At higher surface fraction $\Phi=0.64$, the particles align in ordered
hexagonal lattices (\figref{p120-4}). With only the upward magnetic field
$\bar{B}_z=0.6$ applied, the particles take their upright orientation
$\varphi=0^{\circ}$ and align in disordered arrangements due to the absence of
capillary interactions (\figref{p30-5} and \figref{p120-5}). 
\revisedtext{We perform a Voronoi analysis of the structures (Supplementary Material Fig. S4), 
and demonstrate that the disordered structure in \figref{p120-5}
has a wider dispersion of Voronoi areas as compared to
the locally ordered structure shown in \figref{p120-1}.}
The assembled structures can be tuned by varying the directions of the
magnetic fields (Supplmentary Material Movie S1), which has potential
applications in sensor or display technology. To estimate the time scale of the
assembly and the structural transition, we assume that colloidal particles of
radius $a = 4$ $\mu m$ are adsorbed at a decane-water interface, with a surface
tension $\gamma = 53.2 mN/m$, and the effective viscosity $\mu = 0.91 mPa\cdot
s$\citeref{Fei2018,Adewunmi2019}. Based on our simulation results, the estimated
time scale of structural formation and transition is about $t \sim 1$ $ms$,
which is sufficiently fast to satisfy the requirements of responsive materials for
advanced sensor or display technologies. For a possible experimental realization
of our system, we note that magnetic spherical Janus particles have been
experimentally fabricated\revisedtext{\citeref{Erb2009,Sinn2011,Ren2012,Yan2015}} and investigated at a liquid-liquid
interface\revisedtext{\citeref{Fei2018,Fei2020}}. Such Janus particles can have various amphiphilicities~\cite{Park2011,Kumar2013} and may be stretched
mechanically to form ellipsoidal particles with various aspect
ratios\revisedtext{\citeref{Ho1993,Champion2007, Madivala2009}}. 
\revisedtext{Assuming Janus particles with radius $a = 100 nm$, aspect ratio $\alpha = 3$ and magnetic 
moment $m \approx 4 \times10^{-12} Am^2$ adsorbed at a water-decane interface with a surface tension $\gamma_{ws} = 70
mN/m$, an external magnetic field $H\approx 0.1T$ is able to introduce a magnetic torque larger than the capillary torque and to produce the
various structures observed in our simulations.}


 \begin{figure}[h!]
 \centering
	\includegraphics[width= 0.5\textwidth]{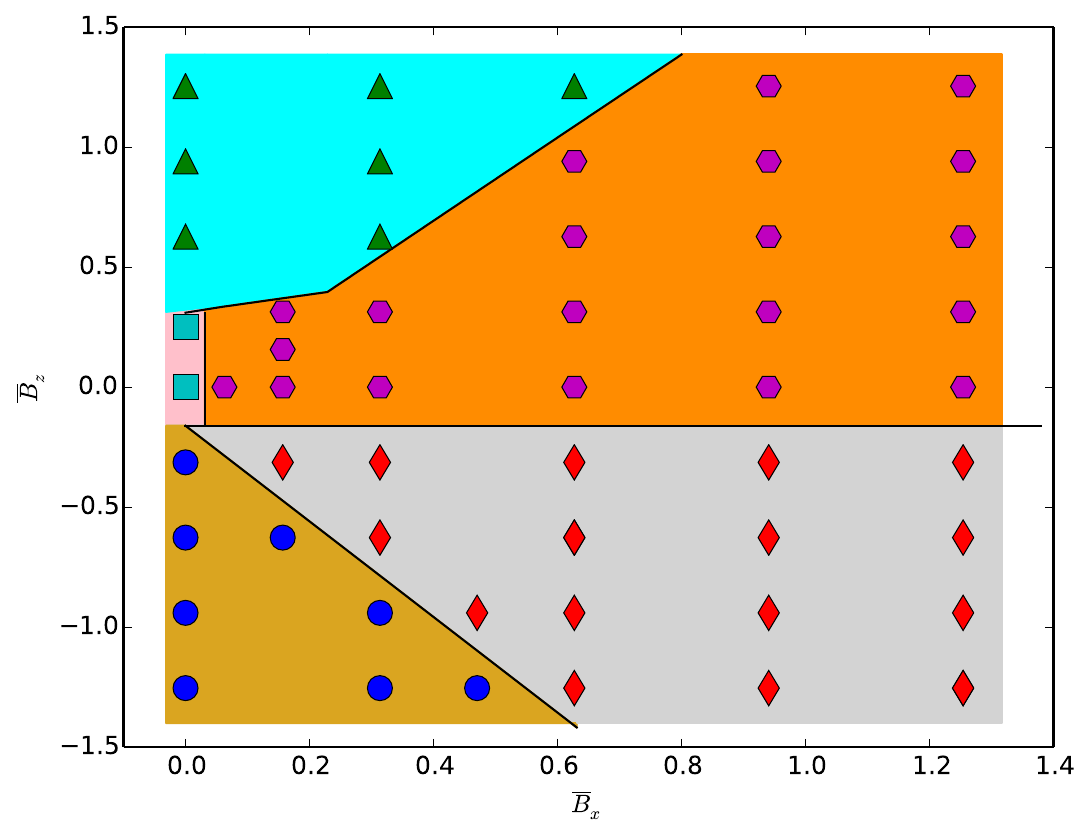}
	\caption{Field-strength phase diagram of assembled phases of Janus
		ellipsoids, showing chains (diamonds), locally-ordered clusters
		(squares), rings (circles), disordered alignments (triangles), and
		hexagonal lattices (pentagons).  The particles have an aspect ratio
		$\alpha=3$ and an amphiphilicity $\beta=21^{\circ}$. The interface
		frction covered by particles is $\Phi=0.62$. }
	\label{fig:phase}
\end{figure}

In~\figref{phase} we construct the phase diagram for assembled structures of
particles as a function of horizontal $\bar{B}_x$ and vertical $\bar{B}_z$
magnetic field-strengths, showing chains (diamonds), locally-ordered clusters
(squares), rings (circles), disordered alignments (triangles), and hexagonal
lattice structures (pentagons). Locally-ordered structures are formed when
external magnetic fields are turned off or only a very weak vertical magnetic
field $-0.1<\bar{B}_z< 0.3$ is applied. Disordered structures occur when the
upward magnetic field is much stronger than the horizontal magnetic field
$\bar{B}_z > 1.7 \bar{B}_x$. In this case, the particles take a small tilt angle
$\varphi<30^{\circ}$, where the deformation of the interface is absent or
negligible.  
The particles form hexagonal lattices once a strong horizontal field is applied
along with an upward magnetic field satisfying $\bar{B}_x/\bar{B}_y>0.6$.
Chains are formed when a strong horizontal magnetic field and a downward
magnetic field is applied, in the range $\bar{B}_x/|\bar{B}_z| > 0.4$.  The
particles assemble into rings when the downward magnetic field is much stronger than the horizontal field $|\bar{B}_z|/\bar{B}_x > 2.5$.

%% file: final.tex
In summary, we demonstrated the controllable self-assembly of ellipsoidal
magnetic Janus particles into clusters, chains, hexagonal lattices and
ring-like structures which can be reconfigured rapidly \revisedtext{by} applying external
magnetic fields. \revisedtext{Our results describe a possible way of creating highly
ordered and, more importantly, tunable structures at fluid-fluid or nematic-fluid~\cite{Wang2019PRL} interfaces} for material assembly.
Possible applications are \revisedtext{omnipresent} not only in the bottom up \revisedtext{fabrication} of
micro/nano-structured surfaces and materials, but also where the ease of
reconfiguration by applying an external magnetic field is of advantage, such as
advanced display, sensor and soft robotics technologies. 
\revisedtext{By applying additionally an oscillating magnetic field, the assembled structures are 
supposed to propel as colloidal swimmers~\cite{Dreyfus2005,Snezhko2011,Lumay2013,Sukhov2019,XieHe2019},
which hold great potential for cargo delivery in biomedical and microfluidic applications.}